\documentstyle[12pt,epsf]{article}
\def\beq{\begin{equation}}
\def\eeq{\end{equation}}
\def\mn{\mu_\nu}
\begin{document}
\begin{titlepage}
\begin{center}
{\Large \bf Theoretical Physics Institute \\
University of Minnesota \\}  \end{center}
\vspace{0.3in}
\begin{flushright}
TPI-MINN-97/15-T \\
UMN-TH-1544-97 \\
May 1997
\end{flushright}
\vspace{0.4in}
\begin{center}
{\large \bf On the Feasibility of Low-Background $Ge-NaI$
Spectrometer for Neutrino Magnetic Moment Measurement}\\[12pt]

{\bf A.G.Beda, E.V.Demidova, A.S.Starostin \\
$SSC$ $RF$  $ITEP,$ $Moscow ~~117259,$ $Russia$ }\\
and\\
{\bf M.B.Voloshin \\
$SSC$ $RF$ $ITEP,$ $Moscow ~~117259,$ $Russia,$ {\bf \it and\\
Theoretical Physics
Institute, University of Minnesota, Minneapolis, MN 55455, USA}  }
\end{center}
\vspace{0.3in}
\begin{abstract}
We analyze the possibility of using a low-background $Ge-NaI$
spectrometer in a reactor experiment for a search for neutrino magnetic
moment down to $3 \times 10^{-11}$ of the electron magneton. The
properties of the so-far existing $Ge$ low-background spectrometers are
discussed and additional sources of background in a reactor experiment
at a small depth are estimated. These estimates place specific
requirements on the design of the spectrometer. The results of
preliminary background measurements at a small depth of 5 m.w.e. with a
dedicated spectrometer built in ITEP are reported.
\end{abstract}

\end{titlepage}

\section{Introduction}

A non-zero magnetic moment of neutrino would be a fundamental physical
quantity, whose implications could lead well beyond the standard picture
of interactions of elementary particles and to non-standard phenomena in
astrophysics.

The recent renewed interest to the problem of the neutrino magnetic
moment (NMM) is in a large part related to the peculiar behavior in time
of the solar neutrino flux measured in the $Cl - Ar$
experiment$^{\cite{davis}}$. One interpretation of this behavior is that
the measured flux anti-correlates in time with the solar
activity$^{\cite{acorr,acorr1}}$. Such an anti-correlation could be
explained$^{\cite{vvo}}$ by interaction of NMM with the
time-dependent magnetic field in the Sun's convection zone. For the
magnetic field in the convection zone in the range $(1 \div 10) \,
KGs$ this mechanism is operative if the magnetic moment of the
electron neutrino is $\mn \sim (10^{-10} \div 10^{-11}) \, \mu_B$,
with $\mu_B=e/(2 \, m_e)$ being the Bohr magneton. Such value of the
NMM would be by several orders of magnitude bigger than the minimal
Standard Model predicts for a massive neutrino$^{\cite{ls}}$:  \beq
\mn = {3\, e \, G_F \over 8 \, \sqrt{2} \, \pi^2} \, m_\nu \approx 3
\times 10^{-19}\, \mu_B \, {m_\nu \over 1 \, eV}~~~.  \label{mstand}
\eeq

However in a number of
extensions of the theory beyond the minimal Standard model (see e.g.
\cite{fy}) one can readily achieve the NMM of the required magnitude,
not necessarily related to the neutrino mass$^{\cite{mv}}$.

The present direct laboratory limit for the NMM for electron
antineutrino is derived from reactor neutrino
experiments$^{\cite{reines,vm,popeko}}$: $\mn \leq (2 \div 4) \times
10^{-10} \mu_B$. More stringent limits on NMM are found from
astrophysical analyses$^{\cite{fuku}}$. Such analyses are typically
based on the fact that at late stages of stellar evolution at densities
above $10^5 \, g \, cm^{-3}$ practically all energy loss by the star is
due to neutrino emission. Thus an NMM would add an electromagnetic
component to neutrino interaction with matter inside the star and by
that significantly modify: the overall rate of cooling, the critical
mass of a He star, the duration and the energy spectrum of the neutrino
burst from a supernova, etc. One should keep in mind however that though
the astrophysical bounds are rather strong: $\mn \leq (0.01 \div 0.1)
\times 10^{-10} \, \mu_B$, they still rely on model dependent
assumptions$^{\cite{sb,bm,raffelt}}$. Therefore it is still very
relevant to improve the sensitivity of direct laboratory measurements of
the NMM, presumably down to $10^{-11} \, \mu_B$.

A laboratory measurement of the NMM is based on its contribution to the
(anti)neutrino - electron scattering. For a non-zero NMM the
differential over the kinetic energy $T$ of the recoil electron cross
section $d \sigma/dT$ is given by the sum of the standard weak
interaction cross section (W) and the electromagnetic (EM)
one\footnote{There is no interference between these two contributions
for a massless neutrino, since the helicity of the final neutrino in a
weak scattering process is opposite to that in a process induced by the
NMM.} (see equations (\ref{sem}) and (\ref{sw}) below). At a small
recoil energy $T \ll E_\nu$ these two components behave in a distinctly
different way: the weak part, $(d\sigma/dT)_{W}$, is practically
constant, while the electromagnetic one, $(d\sigma/dT)_{EM}$ grows as
$1/T$ towards low energies. Therefore for improving the sensitivity to
$\mn$ it is necessary to lower the threshold for detecting the recoil
electrons as far as the background conditions allow. In the experiments
\cite{reines,vm,popeko} the effective range of the detected energy of
the electrons was $T=(1.5-5.2) \, MeV$ because of a sharp increase of
the background at lower energy. In the currently proposed
experiments$^{\cite{bcb,bj}}$, the threshold for registration of the
electrons is expected to be significantly lower: $T_{min} = (0.2-0.6) \,
MeV$. The purpose of this paper is to show that for experiments with the
electron antineutrinos from a reactor a much better sensitivity to $\mn$
per mass of the detector can be achieved with a low background germanium
spectrometer (LBGS) capable of setting the effectively measurable recoil
energy range $T = (3 - 50)\, KeV$. Similar spectrometers are being used
for experiments with double beta decay and in search for weakly
interacting dark matter. A detector of this type dedicated to the search
for NMM is being constructed in ITEP and we report here on its thus far
measured parameters.

In section 2 we discuss general characteristics of $Ge$ spectrometers
that make them suitable for low-background measurements, in particular
for detecting rare events with total energy deposition below $100 \,
KeV$. As existing samples of analogous detectors we discuss the double
beta decay and the dark matter detectors.

In section 3 and 4 is discussed the specifics of low-background
measurements
in a close proximity of a nuclear reactor. We evaluate the cosmic
background conditions and background from inverse beta decay
process with reactor antineutrino.We evaluate also the event rates
from the reactor antineutrinos induced by the weak interactions as
well as by their hypothetical electromagnetic interaction.  Based on
these considerations we give reasoning for the chosen design of the
spectrometer assembly and estimate its
sensitivity to NMM.

In section 5 we estimate the expected effect of the NMM in the
spectrometer and compare it with the realistically achievable
background. This allows to evaluate the possible limits of the
sensitivity in the experiment to the NMM.

In section 6 are described the design and the parameters of the built in
ITEP spectrometer with active $NaI(Tl)$ shielding. We report the results
of tests of the spectrometer in a low-background laboratory in ITEP at
the depth of 5 m.w.e.

\section{Background Characteristics of Germanium Spectrometers}

In detecting very rare events with the rate of less than one event per
day extreme care should be taken with regards to the natural
radioactivity background. This includes using in the construction of the
detector highly purified materials, placing the experiment in special
low-background chambers and applying appropriate selection criteria to
the events.

The advantage of using $Ge$ calorimeters in low-background
measurements is partly based on their high energy resolution $\Delta
E/E = 10^{-3}$ and for the most part on the availability of $Ge$
crystals of extreme purity: the fractions of radioactive
impurities in the $Ge$ crystal are less than $10^{-14}$. A $Ge$
spectrometer can contain either one mono-crystal detector with mass
up to 3 kg, or it can be an assembly of few such detectors with the
total mass (8 - 15) kg.  Since the detectors are quite compact: the
total volume usually does not exceed 4 l, it is possible to use in
its shielding rare and expensive materials: refined electrolytic
copper, highly purified titanium, crystallic silicon, old lead, which is
practically free of the radioactive isotope $^{210}Pb$. As an active
shielding against the charged and the compton components of the
background are used $NaI(Tl)$ scintillators. These special measures
allow to reach very low radiation background  in the double
beta decay and dark matter experiments.

Generally the background spectra of  gamma detectors display a growth
of the compton component of the background towards lower energies. This
is caused by the character of the external radiation
background and by the growth of the detection efficiency at lower
energy of the gamma quanta. In $Ge$ detectors with the surface
contact of the n type this growth of the background becomes slow at
energies (150 - 200) KeV and can turn into a decrease at lower
energies. This is explained by the existence of a passive surface
layer with thickness (0.5 - 1.5) mm in n-p detectors. In other words
the $Ge$ crystal has its own passive shielding that absorbs soft
external gamma radiation.

In the range of energies between 3 KeV and 10 KeV the main sources of
the background are usually the microphonic and electronic noises. The
microphonic noise is caused by vibrations of the detector including the
vibrations induced by the boiling and the turbulent flow of the liquid
nitrogen used as coolant for the detector. The sources of the electronic
noise are the fluctuations of the detector leakage current and the
thermal noise of the FET. The background from the microphonic and the
electronic noises is very efficiently suppressed by the event pulse
shape analysis$^{\cite{gm}}$.

The remaining background at energies (2 - 100) KeV is determined by the
compton continuum and the elastic and inelastic scattering of neutrons
on
$Ge$. The suppression of these components of the background to a great
extent depends on the design of the spectrometer. The existing
installations can be classified into two major types: those with a
passive shielding (PS) and those with an active shielding (AS). An
example of the first type of spectrometer is the installation built
by the USC/PNL collaboration $^{\cite{ba}}$, where two $Ge$ detectors in
copper cryostats are simply placed in a lead housing with 40 cm thick
walls.  However the construction of this spectrometer is based
on the profound expertise of the PNL group in building low-background
detectors using super pure materials. In this installation the inner
10 cm of the passive shielding is made of a very old lead from a
sunken Spanish galleon. Major elements of the construction (the
cryostat, cold finger) are manufactured from electrolytic
copper. In the process of electrolysis from a $CuSO_4$ solution the
copper is purified chemically with high efficiency thus practically
removing the contamination by $U-Th$ admixtures and by cosmogenic
elements produced by the cosmic rays. During the delivery to the
underground site the time the components spent on the surface was
minimized and also the delivery by airplane was excluded, since the
rate of generation of the cosmogenic elements grows by orders of
magnitude at high altitude. As a result after one year of
conditioning of the spectrometer in the underground laboratory the
USC/PNL spectrometer had produced then lowest achieved background count
rate of 0.2 events/(KeV kg year) at the energy about
2 MeV.

An example of an AS installation can be the SB/LBL
spectrometer$^{\cite{ce}}$ for the search for  double beta decay of
$^{76}Ge$. The spectrometer is based on 8 $Ge$ detectors with the
volume 160 - 180 cm$^3$ each, surrounded by an active shielding
consisting of 10 $NaI(Tl)$ scintillator modules ,which in turn is
inside borated polyethylene surrounded by a 20-cm-thick $Pb$
shielding . The veto signals from the scintillators suppressed the
cosmic background. The measurements were taken at the depth of 600
m.w.e. where the cosmic background was still significant. However the
main purpose of the AS was the suppression by a factor 10 - 20 of the
compton component of the radiation background. Thus the scintillator
AS makes the spectrometer less demanding with respect to purity of
the passive shielding and of the rest of the components of the
spectrometer.

Subsequently both spectrometers were used in the search for dark matter.
This had required to lower the threshold down to 4 KeV and to suppress
the background at low energies. As a result the achieved rate of the
background in the energy range (4 - 100) KeV in the USC/PNL and SB/LBL
spectrometers was respectively 0.1 events/(KeV kg day) and 0.4
events/(KeV kg day). If it were possible to have such level of the
background counts in a spectrometer located near a nuclear reactor,
it would allow to significantly improve the sensitivity to the NMM in
comparison with the present bounds$^{\cite{reines,vm,popeko}}$ from
reactor experiments.  However, at much smaller depth and near the
reactor the background conditions are significantly worsened by two
factors: a larger cosmic background and the background from the
reactor itself. Thus far there is no known experience of doing
low-background measurements in such conditions, therefore a
quantitative understanding of the influence of these sources of the
background requires a separate study.

\section{Estimates of the background in a LBGS from the cosmic rays}

The inherent background is practically the same in
AS and PS spectrometers. Therefore the choice of the type of
shielding of a LBGS depends on the ability to suppress the external
background from the cosmic rays and from the working nuclear reactor.
The resultant background is sensitive to a number of factors such as
the exact position of the spectrometer relative to the reactor, the
thickness of the layers of the passive and the active shielding and
the order in which these layers are sandwitched, the presence in the
spectrometer of materials containing hydrogen, etc. Therefore our
comparative estimates of the background and thus of an achievable
sensitivity to the NMM in these two types of spectrometers are made
under the following assumptions:  \begin{itemize} \item{The mass of
the $Ge$ crystal is 2 kg;} \item{The mass of the passive lead shield
is 10 t;} \item{The flux of the antineutrinos from the reactor at the
detector is $2 \times 10^{13} \, \nu/cm^2 \, sec$;} \item{The
installation is located under the reactor at the depth of 20 m.w.e.;}
\item{The thermal neutron background is completely suppressed by the
passive shield of the experimental pavilion;}
\item{The time of experiment consists of 300 days of the reactor on time
and 70 days of the reactor off time.}
\end{itemize}

At a reactor site there is a possibility to place the spectrometer under
the reactor, so that the whole reactor assembly provides a shielding
from the cosmic background with effective thickness of 10 - 40 m.w.e.
At the assumed here effective depth of 20 m.w.e. the hadronic
component of the cosmic background is already very small, and thus
background is dominated by the flux of muons. The counts due to passage
of the muons through the spectrometer and their decay inside the
spectrometer can be easily vetoed out. The most significant remaining
source of the background is the muon capture in the shielding and in the
detector itself due to the reaction$^{\cite{char}}$
\begin{eqnarray}
\mu^{-} + ( Z,\,A ) & \to & ( Z - 1,\, A )^{*}  + \nu_{\mu} \nonumber \\
& \to & (Z-1,\,A-x) + x\,n ~~~,
\label{mucap}
\end{eqnarray}
where $x=0,1,2,\ldots$.

In the capture process a new nucleus is produced
and on the average 1.6 neutrons (in $Pb$) with energy from 6 MeV to few
tens of MeV. The muon capture and the interactions of the secondary
neutrons and the gammas in the detector are separated in time and can
not be vetoed by external counters. The rate of the muon capture events
at 20 m.w.e. is 60 events/kg day$^{\cite{char,gz}}$. Thus in 10 tons
of a passive shielding there will be produced $10^6$ neutrons per day.
The probability for these neutrons to get into the detector depends on
the geometry of the installation and on the presence of a moderator and
absorber layer between the the lead shield and the detector. The muon
capture in the germanium crystal, through the reaction
(\ref{mucap}) produces gallium, whose isotopes undergo beta decay with
the periods $T_{1/2}$ from 30 sec to 14 hours. The results of the
estimates of the background from the $\mu$ capture and at 20 m.w.e. and
of the inherent background of the detector in spectrometers with AS
and with PS are summarized in the Table 1.  \begin{table}
\begin{center}
\begin{tabular}{|c|c|c|}
\hline
Background (events /kg day) & PS spectrometer & AS spectrometer \\
in the energy interval 4 - 50 KeV &           &                 \\
\hline
Inherent background & 5 & 20 \\
\hline
Background from $\mu$ capture & 6 &  \\
in the shielding & 60 & 2\\
\hline
Background from $\mu$ capture &  &  \\
in the $Ge$ detector & 3 & 3 \\
\hline
Total background & 68 & 25 \\
\hline
\end{tabular}
\end{center}
\caption{Summary of the inherent and cosmic background rates in
spectrometers with passive shielding and with active shielding at the
depth of 20 m.w.e. }
\end{table}

One can see from the Table 1 that an AS spectrometer at 20 m.w.e. is
practically insensitive to the cosmic background. This is due to the
action of an internal shielding containing borated polyethylene, which
moderates and captures the neutrons produced in lead. It looks realistic
that a background level of 10 - 15 events/kg day can be achieved in AS
spectrometers by lowering the inherent background of the
spectrometer.  The number used here is based on the measurements with
the SB/LBL installation, which are 10 years old. During this period
there were many changes in practice of low background measurements
and antiradon shielding became the necessary element of such
spectrometers,so it is plausible that a hermetization and filling the
SB/LBL spectrometer with gaseous nitrogen would have significantly
reduced the inherent background.

\section{Background correlated with the reactor activity}

In the proposed here experiment the effect of the NMM is extracted from
comparison of the spectrometer counts with active reactor and with the
reactor shut down. Therefore the background correlated with the reactor
activity requires special attention since it would simulate the effect.
The experience of the reactor neutrino oscillation experiments shows
that the thermal neutrons from the reactor are efficiently absorbed by
the passive shielding of the experimental pavilion. The most serious
source of the background thus would come from the inverse beta decay
process with the reactor antineutrinos:
\beq
A( Z + 1,\, N - 1 ) + \bar \nu_{e} \to A( Z,\, N ) + e^+~~~,
\label{invb}
\eeq
which produces inside the installation unstable isotopes and the
positrons. Most prone to generating this type of background are
materials containing hydrogen, since the cross section of inverse beta
decay on protons is by two orders of magnitude larger than the cross
section of the $\bar \nu_{e} \, e$ scattering. The
only element of the construction containing hydrogen is the borated
polyethylene in the AS spectrometer. We estimate that the reaction
(\ref{invb}) in this part of the spectrometer will contribute $\sim
10^{-3}\, events/kg \, day$ to the background correlated with the
reactor activity. In order to estimate the background from the inverse
beta decay in other elements of construction we evaluated the cross
sections of the reactions $Ge \to Ga$, $Pb \to Tl$, $Cu \to Ni$, $Na \to
Ne$, and $I \to Te$. For all these nuclei the cross section does not
exceed $1.2 \times 10^{-45} \, cm^2$ and the total contribution to the
background is  $\sim 5 \times 10^{-3} \, events/kg \, day$.

\section{The effect of elastic $\bar \nu_e \, e $ scattering}

The differential cross section for the elastic $\bar \nu_e \, e $
scattering associated with the electromagnetic interaction due to the
hypothetical NMM is given by
\beq
{d \sigma_{EM} \over dT}= \left ( {\mn \over \mu_B} \right )^2 \, {\pi
\, \alpha^2 \over m_e^2} \, \left ( {1 \over T} - {1 \over E_\nu} \right
)~~,
\label{sem}
\eeq
while the standard weak cross section is
\beq
{d \sigma_W \over dT} = { G_F^2 \, m_e \over 2 \, \pi} \, \left [
\left ( 1- {T \over E_\nu} \right )^2 \, (1 + 2 \, \sin^2 \theta_W )^2 +
4 \, \sin^4 \theta_W - 2 \, (1+ 2 \, \sin^2 \theta_W ) \, \sin^2
\theta_W \, {m_e T \over E_\nu^2} \right ]~~~,
\label{sw}
\eeq
where $E_\nu$ is the energy of the incident antineutrino, $T$ is the
kinetic energy of the recoil electron, and $\theta_W$ is the Weinberg
angle.

For the range of $T = (3 -50) \, KeV$ one can neglect in
eq.(\ref{sem}) the term $1/E_\nu$ in comparison with $1/T$ and find
the total EM cross section associated with the NMM as \beq
\sigma_{EM} = 7 \times 10^{-45} \, cm^2 \, \left ( {\mn \over 10^{-10}
\, \mu_B} \right )^2~~~.
\label{semn}
\eeq

Thus the excess of the scattering events due to the NMM  in this range
of $T$ and assuming the antineutrino flux $F = 2 \times 10^{13} \,
\nu/cm^2 \, sec$ is estimated as 0.3, 0.5, and 0.8 events/kg day for
$\mn$ respectively equal to $3 \times 10^{-11} \, \mu_B$, $4 \times
10^{-11}\, \mu_B$, and $5 \times 10^{-11} \, \mu_B$. The effect from the
standard weak interaction scattering in the same interval of $T$ is
0.17 events/kg day.

At the threshold energy in the range discussed here one should worry
about the atomic binding effects for the electrons. However using the
appropriate calculations$^{\cite{km}}$ for germanium we find that the
correction due to the binding effects does not exceed 3\% in the
electron energy range (3 - 50) KeV.

Thus the estimates of the cosmic and reactor backgrounds for AS and PS
spectrometers and the estimates of the effects of the electromagnetic
and weak scattering of the reactor antineutrinos on electrons allow us
to conclude the following:
\begin{itemize}
\item{for measurements in the proximity of a reactor one should use a
$Ge$ spectrometer with active shielding, since this allows for a better
suppression of the cosmic background; the background associated with the
reactor activity does not exceed about 3\% of the event rate due to NMM}
\item{for the chosen interval of $T$ and the assumed values of NMM the
number of scattering events due to the NMM ($N_{EM}$) exceeds that due
to the standard weak scattering ($N_W$). For a threshold energy $T >
100 \, KeV$ a reverse relation would hold, which would create a
principal difficulty of extracting $N_{EM}$ from the sum $N_W +
N_{EM}$. Although the weak scattering effect $N_W$ can be calculated,
it contains an uncertainty due to the uncertainty in the energy
spectrum of the reactor antineutrino$^{\cite{hs,bkm}}$. In the energy
range $T = 100 - 1000 \, KeV$ this uncertainty would amount to
(3-7)\%. In the case under consideration ( $\mn > 3 \times 10^{-11}
\, \mu_B, ~~ T = (3 - 50) \, KeV$) the sensitivity of the experiment
is limited only by the statistics.} \end{itemize}

We present in Table 2 the estimates of the achievable sensitivity to the
value of $\mn$ with a 2 kg detector with different assumptions about
the resulting overall background level and for one and two years
duration of the experiment.  We assume that that the measurements are
taken with active reactor for 10 months in a year. One thus sees that
it is realistic that the upper bound on $\mn$ can be lowered down to
(4 - 5)$\times 10^{-11} \, \mu_B$, using a low-background germanium
spectrometer with the mass only about 2 kg.

\begin{table}
\begin{center}
\begin{tabular}{|c|c|c|c|}
\hline
Time of experiment & 25 events/kg day & 15 events/kg day & 10 events/kg
day \\
\hline
1 year & 5.2 & 4.6 & 4.2 \\
\hline
2 years & 4.4 & 3.9 & 3.5 \\
\hline
\end{tabular}
\end{center}
\caption{Achievable 90\% CL upper bounds on the NMM (in units of
$10^{-11} \, \mu_B$ under various assumptions about the overall level of
the background}
\end{table}

\section{Design, construction and parameters of a low-background $Ge -
NaI(Tl)$ spectrometer for NMM measurement}

An installation based on the principles described in the previous
sections and on the experience of participation in low-background double
beta decay experiments of $^{76}Ge\,^{\cite{dbeta}}$ is being
constructed by our group in ITEP. This spectrometer is oriented
towards measurements in proximity of a nuclear reactor with a maximal
suppression of all components of the background.

The low-background spectrometer includes an array of 4 $Ge$ detectors
with total mass above 2 kg inside a $NaI(Tl)$ active shielding. All 4
detectors are mounted in a common cryostat.  All parts of the
cryostat are manufactured from super pure materials:  oxigen-free
copper, titanium, teflon. The shielding assembly has a form of a
cylinder with a central hole. The dimensions of the $NaI$ are : 450 mm
in height, 400 mm in mean diameter with 135 mm diameter central hole.
The assembly consists of 8 light-isolated sections in a common copper
vessel. The ninth $NaI$ crystal closes the central hole from the top
(see Fig.1). As a reflector is used teflon film.  The sections are
monitored by 9 photomultipliers through bent optical guides. This
allows to place the PMs, the voltage dividers and the cables outside
the passive shielding and thus to significantly reduce the inherent
background.  The cryostat and the $NaI$ assembly are surrounded by
passive shielding consisting of 5 cm thick oxigen- free copper, 8 cm
thick layer of borated polyethylene, 15 cm thick layer of lead and
finally one more 8 cm thick layer of  borated polyethylene.  From the
top the spectrometer is covered by an anti-cosmic scintillating
plates with the size $120 \times 120 \times 4 \, cm^3$. For
protection against the radon a double hermetization and a nitrogen
filling of the spectrometer can be engaged. \\

{\bf Current status and results of preliminary tests of the
spectrometer.}

The installation is located in a low-background laboratory in ITEP at
the depth of 5 m.w.e. This allows to suppress the background from a
nearby working accelerator and to significantly attenuate the strong
interacting component of the cosmic background. At present the 4-crystal
detector is being modified for reduction of the microphonic noise, the
anti-cosmic outside scintillator is not installed and the radon
protection is not engaged. The tests were performed with a
one-crystal detector with the volume 106 $cm^3$. Therefore present
results of background measurements are preliminary.

In the actual measurements at the depth of 20 m.w.e. the active
shielding is expected to perform a dual function: to suppress the
compton components in the background spectrum of the $Ge$ detectors and
for subtracting the cosmic induced events. In the tests a conventional
logic was used to process the veto signals. The signal from the $Ge$
detector was fed through a preamplifier and an amplifier to the
analyser. Appropriately shaped signal from the $NaI$ counters was fed to
a summator-extender, after which the veto signals with duration from 10
$\mu$s to 200 $\mu$s was sent to the control input of the analyser. The
threshold of the $NaI$ counters for the gamma quanta was set in the
interval 40 - 50 KeV, since according to previous
measurements$^{\cite{dbeta}}$ the gamma quanta of lower energy do not
reach the counters and are absorbed in the outer inert layers. The
efficiency of the adjustment of the system $Ge$ detector - $NaI$
counters in the veto regime was checked by using a $^{226}Ra$
radioactive source, which was inserted between the cap of the $Ge$
detector and the central $NaI$ crystal. The results of the measurements
were compared with the Monte Carlo simulation using the program GEANT
3.213 .

The discrepancy between the measured and the calculated background
suppression factor does not exceed 15\% over the energy range 50 - 2500
KeV, while the suppression factor itself reaches about 20 at certain
energies. Further modelling using the GEANT 3.213 program revealed
that the background suppression in the range 10 - 150 KeV can be
improved by a factor of two or more by a modification of the detector
construction.

{\bf Background measurements}\\
The study of the background characteristics of the detector was
performed by successively engaging the passive and the active shielding.

The results of the measurements in the energy interval $20 - 5\times
10^4$ KeV are summarized in the Table 3.

\begin{table}
\begin{center}
\begin{tabular}{|c|c|c|c|c|}
\hline
 & E(KeV) & 1 & 2 & 3 \\
\hline
Intensity of the lines & 238.6 ($^{212}Pb$)& 1848 &  --- & 0.18 \\
(events/hour) & 295.2 ($^{214}Pb$) & 890 & --- & 0.21 \\
 & 351.9 ($^{214}Pb$) & 1352 & --- & 0.29 \\
 & 1460.8 ($^{40}K$) & 2944 & --- & 0.12 \\
 & 1764.6 ($^{214}Bi$) & 327 & --- & 0.03 \\
 & 216.5  ($^{208}Ti$) & 492 & --- & 0.02 \\
\hline
Intensity of the  & 20 - 100 & 160 000 & 280 & 5.9 \\
continuum background & 100 - 200 & 220 000 & 320 & 9.9 \\
(events/hour\,100\,KeV) & 200 - 1500 & 18 000 & 68 & 1.5 \\
   & 1500 - 2700 & 800 & 6.7 & 0.10 \\
   & 2700 - 4000 & 6  & 4.8 & 0.02 \\
   & 4000 - 9000 & --- & 2.9 & 0.004 \\
   & $10^4$ - $5 \times 10^4$ & --- & 2.2 & 0.002 \\
\hline
\end{tabular}
\end{center}
\caption{Results of background measurements. 1 - the detector is open, 2
- only the passive shielding engaged, 3 - both the passive and the
active $NaI$ shielding are used.}
\end{table}

The total background count rate in the detector over the energy range 20
- 4000 KeV is 0.01 events/sec, which is already 3 times better than the
 best up-to-date figure for a low-background detector at a small
 depth (15 m.w.e.$^{\cite{eg}}$). The total count rate of the $NaI$
counters when fully covered by the passive shielding is 240
pulse/sec. The efficiency of the suppression of the charged cosmic
component in the energy range 10 - 50 MeV is 99.91\%, which is also
the best figure achieved so far. All the observed lines from the
decay of the elements in the $U-Th$ sequence can be attributed to the
presence in the installation of the radon 220 and 222.

An analysis of these data of the background measurements allows us
to judge on the structure of the background at low energies. Of the
10 events/hour observed in the range 100 - 200 KeV 0.5 events/hour
constitute the inherent radiation background of the spectrometer, 2.0
events/hour are contributed by the radon decay, 0.3 events/hour are
attributed to the inefficiency of the veto system, finally, the dominant
component of the background: 7 events/hour, comes from the secondary
gamma quanta from the hadronic and muonic components of the cosmic
background. These conclusions, derived from the observed spectrum, are
also supported by the results of a simulation with the GEANT 3.213
program.  The results of the measurements of the cosmogenic component
of the background at the depth of 5 m.w.e. do not contradict the
expected suppression at the depth of 20 m.w.e., since at these depths
the rate of the neutron production$^{\cite{gz}}$ (and thus of the
gamma production) by the cosmic rays differs by a factor of more than
20. The study of the structure of the background is ongoing and more
complete and detailed results will be presented elsewhere. \\

{\large \bf Acknowledgements} \

The authors thank M.V.Danilov for support of this work, V.B.Brudanin
for support and many enlightening discussions and A.V.Salamatin for help
and useful advices in electronics.

This work was supported by Russian Fund on Fundamental Nuclear
Physics,grant 1.3.1-04 . The work of MBV is supported, in part, by
the DOE grant DE-AC02-83ER40105.

\newpage
\thispagestyle{empty}
\begin{figure}
\epsfysize=8.0in
\epsfbox[90 120 526 727]{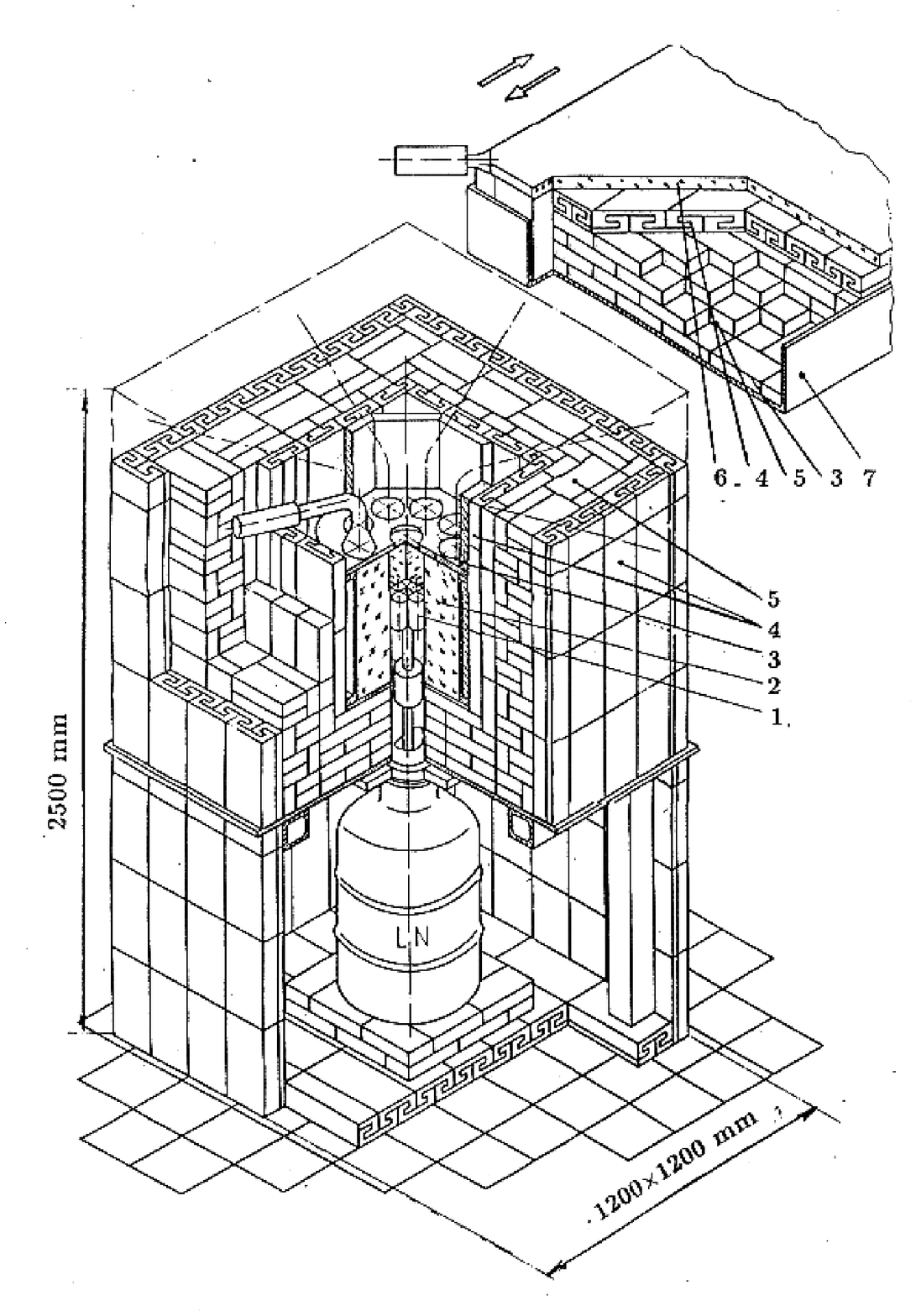}
\caption{The low background $Ge-NaI(Tl)$ spectrometer for NMM measurement. 1 - $Ge$ crystal, 2 - $NaI(Tl)$, 3 - oxigen-free copper, 4 - borated polyethylene, 5 - lead, 6 - anti-cosmic scintillator, 7 - moveable top shielding assembly. }
\end{figure}
\end{document}